%% file: mobisys21-sleep.tex
\documentclass[acmlarge]{acmart}

\newcommand{\transhealth}[1]{\textcolor{black}{#1}}
\newcommand{\revise}[1]{\textcolor{black}{#1}}
\newcommand{\wisleeps}{{\em WiSleep\ }}
\newcommand{\wisleep}{{\em WiSleep}}

\newcommand{\etals}{{\em et al. }}

\usepackage{pbox}
\usepackage{array}
\usepackage{subcaption}
\usepackage{makecell}
\usepackage{multirow}
\usepackage{color, colortbl}
\definecolor{LightCyan}{rgb}{0.88,1,1}

\AtBeginDocument{%
  \providecommand\BibTeX{{%
    \normalfont B\kern-0.5em{\scshape i\kern-0.25em b}\kern-0.8em\TeX}}}

\copyrightyear{2022} 
\acmYear{2022} 
\setcopyright{acmlicensed}\acmConference[X]{X}{X}{X}
\acmBooktitle{X}
\acmPrice{15.00}
\acmDOI{X}
\acmISBN{X}

\begin{document}

\title{\transhealth{WiSleep: Inferring Sleep Duration at Scale Using Passive WiFi Sensing}}


\author{Priyanka Mary Mammen}
\affiliation{%
  \institution{University of Massachusetts Amherst}
  \country{USA}}
\email{pmammen@cs.umass.edu}

\author{Camellia Zakaria}
\affiliation{%
  \institution{University of Massachusetts Amherst}
  \country{USA}}
\email{nurcamellia@cs.umass.edu}

\author{Tergel Molom-Ochir}
\affiliation{%
  \institution{University of Massachusetts Amherst}
  \country{USA}}
\email{tmolomochir@umass.edu}

\author{Amee Trivedi}
\affiliation{%
  \institution{University of Massachusetts Amherst}
  \country{USA}}
\email{amee@cs.umass.edu}

\author{Prashant Shenoy}
\affiliation{%
  \institution{University of Massachusetts Amherst}
  \country{USA}}
\email{shenoy@cs.umass.edu}

\author{Rajesh Balan}
\affiliation{%
  \institution{Singapore Management University}
  \country{Singapore}}
\email{rajesh@smu.edu.sg}

\renewcommand{\shortauthors}{Mammen, Zakaria, Molom-Ochir, Trivedi, Shenoy and Balan, et al.}

\input{abstract}



\maketitle
\input{Introduction}
\input{Background}
\input{Approach}

\input{System}
\input{Validation}
\input{Scalability}

\input{Casestudy}
\input{Discussion}

\input{Conclusions}


\bibliographystyle{ACM-Reference-Format}
\bibliography{mobisys21-sleep}

\end{document}

%% file: abstract.tex
\begin{abstract}
Sleep deprivation is a public health concern that significantly impacts one's well-being and performance. Sleep is an intimate experience, and state-of-the-art sleep monitoring solutions are highly-personalized to individual users. With a motivation to expand sleep monitoring capabilities at a large scale and contribute sleep data to public health understanding, we present \wisleep, a \transhealth{system for inferring sleep duration} using smartphone network connections that are passively sensed from WiFi infrastructure. We propose an unsupervised ensemble model of Bayesian change point detection, \transhealth{validating it over a user study among 20 students living in campus dormitories and a private home.} Our results find \wisleeps outperforming prior techniques for users with irregular sleep patterns while yielding \transhealth{ an average 88.50\% accuracy within 60 minutes sleep time error and 39 minutes wake-up time error}. This is comparable to client-side methods, albeit utilizing coarse-grained information. \transhealth{Additionally, we utilize our approach to predict sleep and wake-up times from a user study of more than 1000 student users, demonstrating results similar to prior findings on students' sleep patterns.} Finally, we show that \wisleeps can process data from twenty thousand users on a single commodity server, allowing it to scale to large campus populations with low server requirements. 
\end{abstract}

%% file: Introduction.tex
\section{Introduction}
\label{sec:introduction}
Sleep is a vital activity that significantly impacts human well-being, productivity and performance \cite{rosekind2010cost}. Prior research has shown that 30\% of the adult population does not get enough sleep, with many adults sleeping less than 7 hours per day \cite{centers2009perceived,krueger2009sleep}. Both work-related stress and the increasing use of mobile devices throughout the day, particularly in the evenings, have increased sleep disorders \cite{thomee2011mobile}. The repercussions of sleep deprivation leading to serious health consequences such as heart disease, stroke, and depression \cite{altevogt2006sleep,perry2013raising} \transhealth{has} become a public health burden. The American Academy of Pediatrics confirms sleep deprivation as a public health epidemic, especially among students \cite{perry2013raising,adolescent2014school}.

Sleep is an intimate experience; hence many sleep monitoring technologies are highly personalized for individual use. Monitoring data sources specific to sleep are challenging to acquire for public health understanding and benefits \cite{perry2013raising}. Such information could benefit professional health administrators to keep abreast of a community's needs and well-being. In particular, college students \transhealth{residing} in campus dormitories make an insightful study population of irregular sleepers due to overwhelming academic demands. Further, many college campuses, such as in the United States, are known for their social events during the semester. The active party culture exacerbates bad sleeping habits among students \cite{vail2009relationship}. These irregular habits can significantly and negatively impact students' concentration and academic performance \cite{maheshwari2019impact}.

Numerous solutions have emerged for sleep monitoring. Polysomnography is a gold standard in medical research \cite{ruehland20112007} practical for short-term monitoring. The consumer market witnessed growing alternatives for sleep trackers using accelerometers, heart rate sensors \cite{fitbit,applewatch}, and soon, research prototypes of in-ear devices \cite{Nguyen16} will follow. However, there remains a general reluctance to use wearable while sleeping \cite{rahman2015dopplesleep}, making it challenging to support long-term monitoring. Contactless methods utilizing doppler radar and RF signals have been proposed \cite{rahman2015dopplesleep,hsu2017zero}, but they require specific instrumentation in building infrastructures. Much prior work includes using smartphone sensors, such as microphones, cameras, phone activity, and screen usage  \cite{gu2015sleep,min2014toss,hao2013isleep,Cuttone17,Zhenyu13}. These client-side approaches require direct sensing of users' devices. Unfortunately, such methods cannot be easily scaled to large groups of users and would eventually face pushback over privacy concerns.

Our work focuses on \emph{the challenge of developing a  sleep \transhealth{analytics system} at scale for public health benefits while still rendering personal wellness goals}. With growing efforts in monitoring students' mental health and well-being using sensing technologies \cite{wang2018tracking,zakaria2019stressmon}, our work's broader goal is to incorporate \transhealth{capabilities that can support sleep health assessment} into these efforts and offer a holistic community well-being service. 

In this paper, we present \wisleep\footnote{Pronounced \emph{why-sleep}, an apt name for a system for estimating sleep times among students.}, a \transhealth{system for inferring sleep durations using coarse-grained WiFi events that are passively sensed from the WiFi infrastructure. \wisleeps adopts a network-centric approach, which does not require on-device data collection or apps to be downloaded on the phone. Without changing users' behavior on how they use their smartphones, it leverages the network's view of the user's device as a proxy of their activity and location. Specifically, when a user uses their smartphone, it generates WiFi network events such as connecting to an access point, indicative of user activity. Conversely, when a user sleeps, their smartphone generates little to no WiFi network activity. \wisleeps analyses WiFi logs of network events generated by a user's smartphone to detect prolonged periods of inactivity. It infers users' sleep periods from such inactive periods. We show that the passive observations of declining association type network events from users' smartphones can infer their sleep.}

Our key design goal is \textit{scalability} to promote and support adoptions at population-scale, for example, among large groups of students on college campuses. Firstly, choosing a network-side approach helps scale our technology rapidly to every device (and, by proxy every, user of these devices) as soon as they are connected to the WiFi network, without requiring active client participation. Second, our approach is sufficiently ready for immediate deployment without requiring additional hardware installation to the WiFi infrastructure. Third, although we focus on college campuses as our target community, we also demonstrate that our approach is general and work for users in private homes. This paper makes the following contributions:

\begin{enumerate}
    \item  We present a model based on Bayesian Change Point detection to predict sleep periods utilizing coarse-grained network events of smartphones. Our user study involves predicting sleep and wake-up time of more than 1000 students. Validation of our model using ground truth data from \transhealth{20} users living on campus and a private house owner yielded \transhealth{an average accuracy of 88.50\%, 78.39\% precision, 86.90\% recall, and 0.84 F-score, within 60 and 39 minutes of sleep and wake-up time error.} We find our model robust to noisy data and detecting irregular sleep patterns, which are common among our targeted population. This unsupervised method requires no prior training data, enabling a scalable approach to large user populations.
    \item We investigate practical challenges by addressing confounding factors \transhealth{related to device behaviors, such as} WiFi AP ping-pong effects and background network activity on smartphones. We clarify how device inactivity during the day does not affect our prediction results. Further, we demonstrate \wisleeps to scale from 15 to ten thousand users using one server, processing one user in approximately 4 seconds.
    
    \item We conduct two case studies to show the value of our analytics platform for population and personal use. The first analyzes \transhealth{anonymized data from} 1000 on-campus student residents over a week, informing different student groups' sleep behaviors. These findings can supplement public health's understanding of sleep-related problems. A second longitudinal analysis of students over one semester can help individuals understand their sleeping habits by the hour of day and day of the week.
\end{enumerate}

%% file: Background.tex
\section{Related Work}
\label{sec:intro}
In this section, we present the prior efforts related to detection and monitoring solutions.

\begin{table*} [h!]
\centering
\scalebox{1}{
\begin{tabular}{ |l|c|c|c|c| } \hline
\textbf{Approach} & \textbf{Sleep Ability} & \textbf{Contactless} & \textbf{Supervised} & \textbf{Deployment} \\ \hline
Doppler$[34]$,RF$[20,26]$ & duration,quality & yes & yes & building \\ \hline
In-ear$[30]$ & duration/quality & no & yes & wearable \\ \hline
Phone activity$[17,18,29]$ & duration & yes & yes & smartphone \\ \hline
Screen activity$[11]$ & duration & yes & no & smartphone \\ \hline
\wisleep & duration & yes & no & WiFi \\ \hline
\end{tabular}}
\caption{A comparison of prior approaches.}
\label{tbl:prior}
\end{table*}

Many of these efforts are built on IoT and wearable devices to address individual users' sensing requirements. These devices are increasingly accepted for everyday use, but they are not as ubiquitous as other mobile devices such as smartphones. This is an important consideration as \textit{scalability} is a key capability of our system. Our focus is to facilitate population-scale sensing while maintaining personal use. In this work, we consider students residing on campus as our large population sample. In what follows, we describe established approaches and how our technique aims to address their shortfalls. These works are summarized in Table \ref{tbl:prior}.\\

\noindent\textbf{Wearables}
Sleep monitoring over long periods has become feasible due to the availability of wearables. Consumer trackers such as Fitbit \cite{fitbit} leverage accelerometer or heart rate data. Researchers have explored novel methods such as in-ear wearable sensors to precisely monitor sleep quality and duration \cite{Nguyen16}. By design, wearables are appropriate for individual monitoring. They can support population-scale monitoring, but all users must wear such devices and transmit the sensed data to the cloud for large-scale analysis. Fitness trackers are still not ubiquitous despite their increased popularity and impose a deployment cost for large users. More importantly, many users remain reluctant to wear an on-body device while sleeping \cite{rahman2015dopplesleep}.\\

\noindent\textbf{Contactless Techniques}
In contrast, contactless sleep monitoring overcomes adoption pushbacks by installing sensors in the environment (e.g., wall sensors). These efforts include Doppler radar or RF signals to sense sleep patterns~\cite{rahman2015dopplesleep,hsu2017zero,liu2015tracking}. While such techniques show significant promise, they incur a higher cost for population-scale sensing due to the need to deploy instrumentation in buildings (e.g., all dorm rooms in a college campus).\\

\noindent\textbf{Mobile Sensing}
The ubiquity of smartphones motivates many researchers to use phone-based sensors as sleep trackers. These works included microphones, cameras and phone activity logs \cite{gu2015sleep,min2014toss,hao2013isleep,ren2015fine,Cuttone17,Zhenyu13}. Others have shown monitoring screen activities can be an effective method to infer sleep and wake \cite{Cuttone17,Abdullah14} due to the strong correlation between (lack of) phone activity and user's sleep. Such a method does not incur any hardware deployment costs due to the ubiquity of smartphones. However, client-side smartphone-based methods face different challenges to scale up to a large number of users. First, they require dedicated apps, which can be a hurdle at population scales\footnote{Mandating an app to be installed by all users is challenging to enforce, as has been noted from the poor uptake of COVID-19 contact tracing apps even in places like Singapore where they were strongly recommended \cite{singapore_contact}}. Second, longitudinal monitoring can be an issue when users change or upgrade phones. This practice is typical among tech-savvy student users, and device changes impose re-installation overheads, which is hard to automate. \\

\noindent\textbf{Detection Mechanism}
All of the above methods can be classified as being supervised or unsupervised. Supervised approaches require training data to build detection models. Since collecting large amounts of training data is challenging, a supervised approach is generally harder to scale. In contrast, unsupervised approaches, such as Bayesian methods, do not need any training data and are easier to deploy at a population scale. For example, Khadiri \etals and Cuttone \etals employed unsupervised Bayesian inference to infer sleep periods using different types of sensors \cite{Khadiri18,Cuttone17}. Similarly, an unsupervised approach is best in our case to build a detection model without worrying about training models.

\section{Using Passive WiFi Sensing}
\label{sec:motivation}
The shortcomings of prior work inform our decision to leverage a network-based sensing approach. In what follows, we justify our considerations to adopt WiFi-based sensing.

Prior works have employed WiFi logs to predict other types of human behavior, such as respiration \cite{khan2017deep}, social interaction, and routined activities \cite{kalogianni2015passive,hong2016socialprobe,zakaria2019stressmon}. Using the same modality, our work investigates its use in \transhealth{estimating sleep duration}. As humans grow increasingly reliant on their smartphones \cite{van2015modeling}, much of the common online access such as video streaming, mobile gaming, and virtual communication demand low latency and high bandwidth networks that WiFi can offer \cite{sommers2012cell}. Because of this, WiFi is more preferred as home network solution, running more efficiently in the long run than relying on cellular networks \cite{ware2018large}.

\subsection{\transhealth{Motivation}}
\label{sec:motivation}
\transhealth{Our work focuses on using the network events generated by a user's smartphone when it connects to the WiFi network to proxy user activity. Specifically, our approach involves observing reports of WiFi network activity to infer if the user is awake or asleep. When a user is awake and uses their smartphone, the device generates WiFi network events such as connecting to an access point. Conversely, the device generates little to no WiFi network activity when a user sleeps.} 

We hypothesize that \textit{network activity from a student's phone is strongly correlated to the user's activity and awake state, thus simply observing these network activities through coarse-grained AP association events, is sufficient to infer sleep periods}. Similar to techniques relying on phone's screen activity, we expect long periods of low network activity are correlated to sleep periods. 

\begin{figure}[h!]
	\centering
	\includegraphics[width=.5\textwidth]{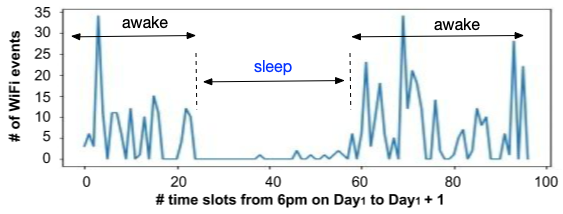}
	\caption{Smartphone network events over 24 hours, with low event rate corresponding to sleep.}
	\label{fig:poisson}
\end{figure}

\transhealth{Consider the time-series example of a user's smartphone network events throughout a 24-hour period to understand why this is feasible. Figure~\ref{fig:poisson} illustrates a user's smartphone network events over a 15-minutes interval from Day 1, 6 pm to Day 2, 6 pm. When a user establishes a WiFi connection on their phone for online communication,} the device will connect to a nearby AP, generating network events. The device will periodically re-associate to stay connected to the best AP for as long as the user needs the connection, thus, triggering a sequence of association and disassociation events. The device eventually falls into a power-saving state when the user stops interacting with it. Periodically, the device `wakes up' (e.g., every 15 to 30 minutes) and performs a network scan that triggers a re-association. 
The fluctuations in network events help us predict the user's activity and state. The main challenge is in determining which period of low network activity should accurately infer a user as (actually) sleeping.

\subsection{Passive WiFi Sensing: A Network-centric Approach}
\label{sec:networkcentric}
\transhealth{Observations of a phone's network activity can be done on the client-side (i.e., on the user's phone) or from the network-side since networks capture the phone's network traffic and events. Our approach is entirely network-based, meaning that it solely relies on the network observation of users' devices when connected to WiFi as a proxy of user behavior. Doing so does not require users to change their natural behavior of using a wearable device for tracking or installing a new app on their smartphone. Specifically, we employ the output logs (i.e., called syslog) generated from the WiFi AP that the device is connected to. These network events are specifically two events types -- \emph{association} and \emph{disassociation} events. Briefly, when a phone connects to a WiFi AP, it is said to \emph{associate}, generating \emph{association} events. When it disconnects, it \emph{disassociates} and produces \emph{disassociation} events. Our approach examines these events, which are recorded in WiFi logs.}  

\transhealth{Many enterprise WiFi and home WiFi routers provide logging capabilities of coarse-grained network events for the network's performance and security monitoring; in such cases,  our approach can utilize these logs without the need to collect any additional data. Besides doing away with a dedicated mobile app, the approach is impervious to changes (e.g., when users upgrade their devices to newer models). By utilizing existing WiFi networks and ubiquitous smartphones, we also avoid additional deployment costs. Despite its benefits, it is important to note that WiFi network events are coarse-grained and inherently noisy \cite{trivedi2020wifitrace}, hence can increase false positives from inaccuracies. The following section describes how such inaccuracies can arise from various user behaviors. Nevertheless, to the best of our knowledge, our proposed technique is the first to infer sleep periods using a network-centric approach.}

\subsection{Challenges} 
\label{sec:designRationale}
\transhealth{Any approach that uses a phone's activity as a proxy for a user's activity to infer sleep and wake periods needs to handle many challenges. Both users' behavior and device behaviors can introduce errors. From the user's perspective, it is possible that errors are introduced as a result of:}
\begin{enumerate}
    \item \transhealth{\textbf{Sleep onset latency:} a user may not immediately fall asleep upon putting their phones away at night, much more than they would from waking up in the morning (e.g., respond to alarm).}
    \item \transhealth{\textbf{Delayed morning smartphone routine:} a user may not check their phone as soon as they wake up. }
    \item \transhealth{\textbf{Intermittent wake-ups:} a user briefly wakes up during their nocturnal sleep but may not use their phone.}
\end{enumerate}

\transhealth{From the device's perspective, it is possible that errors are introduced as a result of:}
\begin{enumerate}
    \item \transhealth{\textbf{Ping-pong effects:} the smartphone automatically switches its network connection between nearby APs for the most optimal connection; thus this network activity incorrectly infers user behavior.}
    \item \transhealth{\textbf{Background activities:} the smartphone performs activities independent of the user, such as installing a software update, receiving emails and messages, incorrectly inferring user behavior.} 
\end{enumerate}

\begin{figure} [t]
	\centering
	\includegraphics[width=.5\textwidth]{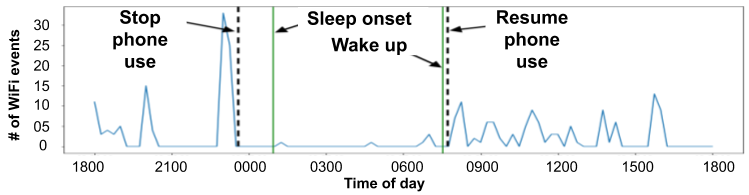}
	\caption{Potential sensing errors between ceasing/resuming phone activity and sleep/wake onset.}
	\label{fig:wifiStopResume}
	\vspace{-4mm}
\end{figure}

\transhealth{It is important to emphasize that our approach aims to infer users' sleep duration -- and not detect the nuances in sleep characteristics \cite{rahman2015dopplesleep,Nguyen16} that prior work is set to achieve.} For example, as illustrated in Figure \ref{fig:wifiStopResume}, ceasing phone activities before sleep time does not immediately translate to sleep onset, as users may take some time to fall asleep. For such reasons, it is \textit{difficult to tackle our work as a simple binary classification problem where the longest sequence of low activity periods over a day is determined as the sleeping period.}

The rest of the paper describes how challenges \transhealth{related to device behaviors} are addressed and can support compelling utility at a population scale. We demonstrate through our primary use-case, focusing on inferring sleep among the student population and its utility for the student's health and well-being services. A supplementary study on a private home shows how our approach \transhealth{can be applied} for personal use.



%% file: Approach.tex
\section{Bayesian Sleep Inference}
\label{sec:approach}
Our detection mechanism is an ensemble method based on Bayesian change point model. In what follows, we describe the problem statement to build our model.\\


\noindent \textbf{Problem Statement}
As stated in Section \ref{sec:motivation}, we assume the primary user's device to be smartphones. Consider an enterprise WiFi network deployed in a university campus with \textit{M} access points and \textit{N} users. We model each user as being in one of two states: \textit{awake} or \textit{sleeping}. When the user is `awake', they can either be mobile (moving from one location to another) or localized at a given area and assumed to use their phone from time to time frequently. In either case, the phone generates AP association and disassociation events logged by the network (we explain this in detail in Section \ref{sec:datacollectionengine}). 
With a 24-hour trace of time-stamped WiFi events, we use this trace to compute the \textit{rate of network events}; we divide the 24-hour period into time slots and count the number of events in each slot. Let $w_t$ denote WiFi event rate seen at time $t$ and let $b$ denote the slot size (we choose a default slot size of $b$=15min, yielding 96 slots per day). Given a time series of event rates $w_t$, our problem is to estimate the sleep onset time, $T_{sleep}$, and the wake-up time, $T_{awake}$ for the user.

\subsection{Bayesian Change Point Detection}
We estimate the sleep and wake-up times from WiFi events based on Bayesian Change Point detection, well established to detect significant changes in time-series data and have been widely used for anomaly detection. As illustrated in Figure \ref{fig:poisson}, we must as accurately detect a significant drop in the phone's network activity that occurs at sleep time and a corresponding rise that occurs upon a wake time. Hence, $T_{sleep}$ and $T_{awake}$ are significant change points that we must detect in our time series data $w_t$, based on Bayesian inference of change points.


We model $w_t$ as a Poisson process (i.e, a time series of event rates in a time slot is Poisson), where $\lambda$ is the mean of the distribution.
 \[
 P(w)= Poisson(w, \lambda) = \frac{\lambda^{w} e^{-\lambda}} {w!}
 \]

Since the mean event rate $\lambda$ drops at sleep onset time $T_{sleep}$ and rises at wake-up time $T_{awake}$, therefore $\lambda_{sleep}$ and $\lambda_{awake}$ denote the mean event rate when a user is asleep and awake. 
\begin{equation}
\lambda= \left \{
\begin{aligned}\lambda_{sleep}, && \text{if}\ T_{sleep} \leq t < T_{awake} \\
\lambda_{awake}, && \text{Otherwise}
\end{aligned} \right.
\end{equation}

Since the mean event rate $\lambda_{awake}$ is high when the user is awake and the event rate $\lambda_{sleep}$ is low when asleep (see Figure \ref{fig:poisson}), we assume that $\lambda$ follows a gamma distribution with the following density function.
\[ \Gamma(\lambda, a, b)= \frac{1}{\Gamma(a)}b^{a}\lambda^{a-1}exp(-b\lambda)
\]
 
Given these assumptions, we need to detect two change points $T_{sleep}$ and $T_{awake}$ when the event rate in the time series transitions from $\lambda_{awake}$ to $\lambda_{sleep}$ and vice versa. Bayesian change point detection involves finding the posterior distribution of the change points for different values of $t$ and maximizing it to derive the Maximum A Posterior Estimates (MAP). 
 This is done by using a Metropolis-Hastings algorithm \cite{chib1995understanding} to estimate these parameters for each value of $t$ and choosing the $t$ that corresponds to MAP as the change point. As in any Bayesian approach, we need to assign priors to the model parameters (i.e, $\lambda_{sleep}$, $\lambda_{awake}$, $T_{sleep}$, $T_{awake}$ ) and then use Metropolis sampling to derive posterior conditional distribution of each parameter from its joint distribution. As noted earlier, the value of $t$ where the distribution is maximized (MAP) represents the change point $T_{sleep}$ (and $T_{awake}$).

%
%

\subsection{Ensemble Model for Sleep Inference}
\label{sec:ensembleModel}
The need for our Bayesian approach to be robust to noisy WiFi data and irregular sleep patterns (see Section \ref{sec:designRationale}) makes it challenging to build a model with strong priors -- consequently, models with weak (or non-informative) priors impact model accuracy. Accordingly, we employ an ensemble method comprising three separate models, each with priors suitable for different scenarios, and finally, apply Bayesian Model Averaging (BMA) \cite{Fragoso_2017} to derive a combined estimate. The composition of our ensemble model is:

\subsubsection*{Model 1) Bayesian Model with Location-based Non-informative Prior} assumes that the sleep periods occur in one or a small subset of locations, such as a dorm room. The location information is inferred directly from the AP placements without localizing the device itself. Priors for a particular day are chosen based on the times spent at these locations. This model is useful for users who have irregular sleep hours but consistent sleep locations. Such location based priors avoid choosing time periods spent outside the dorm areas for possible sleep periods.

To specify the prior for a specific day, we assume the mapping of all campus APs to their building locations are known a priori and only consider a subset of APs located in the residential dorms. For every user's 24-hour WiFi trace, we determine the longest duration spent in a dorm building (based on network activity observed by the dorm APs). Note, however, this assumption ignores sleeping behaviors outside the dorm area.

Let $[T_{start}, T_{end}]$ denote the time-interval spent in dorm areas, $k$ hours as the minimum sleep duration (e.g., $k=3$ hours is equivalent to 12 time slots of 15-min intervals). Since sleep patterns can be irregular, we assume $T_{sleep}$ and $T_{awake}$ are uniformly distributed within $[T_{start}, T_{end}]$. Hence, the model priors are given as:
 \[ T_{sleep} \sim DiscreteUniform(T_{start},T_{end}-12)\]
 \[ T_{awake} \sim DiscreteUniform(T_{start}+12,T_{end})\]

The event rate while awake is assumed to be 2.5 events/bin yielding a prior:
 \[\lambda_{awake} \sim Gamma(2.5,1)\]
    
The event rate while sleeping is assumed to be a low non-zero rate:
 \[\lambda_{sleep} \sim Gamma(1,1)\]

\subsubsection*{Model 2) Bayesian Model with Normal Prior} assumes that the sleep onset and wake-up times are normally distributed (rather than uniformly distributed as in the previous model), thus suited for users with regular sleep and wake-up times.

Let $T_{start}$ and $T_{end}$ denote the start and end times of their daily sleep period. $T_{start}$ and $T_{end}$ are normally distributed with a standard deviation $\sigma$. Assume that a student goes to sleep at 12:00 am and wakes up at 8:00 am the next day, with a standard deviation of 3 hours ($T_{start}$=12am, $T_{end}$=8am, $\sigma=$3). The priors for $\lambda_{sleep}$ and $\lambda_{awake}$ are the same for all models. Hence, the model priors are given as:
 \[T_{sleep} \sim Normal(T_{start},12)\]
 \[T_{awake} \sim Normal(T_{end},12)\]
 
\subsubsection*{Model 3) Bayesian Model with Hierarchical Prior} is useful when sleep behavior changes based on the day's events, resulting in varying standard deviation.

Let $T_{start}$ and $T_{end}$ denote the start and end times of a sleeping period, normally distributed as per Model 2 ($T_{start}$=12am, $T_{end}$=8am). As sleep behavior varies based on the day's events, $T_{sleep}$ and $T_{awake}$ can be derived by adding hyper-priors $\alpha_{t}$, $\beta_{t}$ and $\tau_{t}$ to the normal priors. We set the hyper-priors to a non-informative distribution since we have no strong knowledge about them. The priors for $\lambda_{sleep}$ and $\lambda_{awake}$ are the same for all models. Hence, the model priors are given as:
 \[\alpha_{t}\sim Exponential(1) \]
 \[\beta_{t}\sim Exponential(1) \]
 \[\tau_{t}\sim Gamma(\alpha_{t},\beta_{t}) \]
 \[T_{sleep} \sim Normal(T_{start},\tau_{t})\]
 \[T_{awake}\sim Normal(T_{end},\tau_{t})\]
 
Once all models are utilized for change point detection, these results are averaged using Bayesian Model Averaging. All models are weighted using a marginal likelihood where the weights are sensitive to the prior distribution. We generate the weights from the posterior distribution of these models using the Watanabe-Akaike Information Criteria (WAIC) \cite{Watanabe13}. WAIC relies on the complete posterior distribution rather than on a single point estimate, making it a more robust approach for generating a combined estimate from the ensemble predictions.

%% file: System.tex
\section{\wisleeps System Overview}
\label{sec:impl}

\begin{figure}[h!]
    \centering
    \includegraphics[width=.9\textwidth]{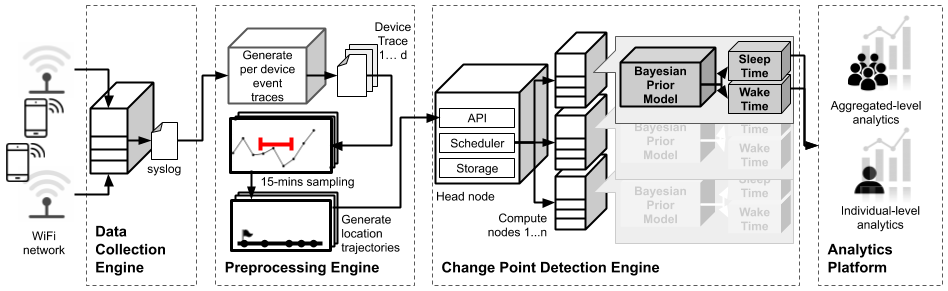}
    \caption{System components of \wisleep.}
    \label{fig:systemoverview}
    \vspace{-4mm}
\end{figure}

\transhealth{We now describe each system component of \wisleep, as illustrated in Figure \ref{fig:systemoverview}. We have built a prototype of our system and  deployed it in our university campus and a home setting.}

\subsection{Data collection Engine}
\label{sec:datacollectionengine}
\transhealth{\wisleeps utilizes network events generated by a phone when it connects to the WiFi network to proxy user activity. While leveraging various network event types presents more information of the user, \wisleeps only uses coarse-grained information. In particular, it uses two types of events \emph{association} and \emph{disassociation}. To obtain these events, we collect event logs that are generated by all modern WiFi APs (e.g., \emph{syslog}) \cite{aruba-syslog}. A residential setting typically consists of a single AP or a WiFi mesh that logs the connection events of devices, and event log collection is possible in any programmable router. In an enterprise setting such as campus, a network of APs does the logging. As an example, our campus deployment consists of 5500 HP/Aruba wireless APs, managed by seven wireless controllers. The syslog data is ultimately sent to a central syslog server for data aggregation of multiple IT systems and network components. As will be explained in Section \ref{sec:datauser}, collection of WiFi network logs were with the approval agreement from the university's IT department and home owner. Each AP keeps a log comprising a sequence of timestamped events in the format:} 

\begin{verbatim}
hh:mm:ss <controller_name> <process_id> <event_subtype> <MAC_addr> <event_body>
\end{verbatim}

\transhealth{While there are many fields to this log, the most relevant to our work are \emph{timestamp}, \emph{controller\_name}, \\ \emph{event\_subtype} and \emph{MAC\_addr}. A device can be identified through its MAC address (\emph{MAC\_addr}). When a device connects and disconnects to an AP, the AP will log an event. It is in this \emph{event\_subtype} field that we can distinguish different event types, particularly \textit{association}, \textit{disassociation} events. Specifically, when a user's device connects to WiFi, the connection is established with the nearest WiFi AP from where the user is located, generating \textit{association} type events. The user's device stays connected to the AP for as long as they utilize the network and remain in that location. When they move to another location, the WiFi connection switches to the next nearest AP to where the user is now situated. Accordingly, \textit{disassociation} and  \textit{re-association} events will be generated when the user's device moves out of range and reconnects to the network. Similarly, when a user's device becomes inactive, the AP will log a \textit{disassociation} event. The \emph{controller\_name} gives us information on where the user is located based on the nearest AP their device is connected to. It is important to note that throughout the whole time, the user is assumed to maintain the same network connection to the campus WiFi, but is only switching APs as they move.}

\subsubsection*{Scalability:}
\wisleeps has no specific data collection scalability challenges to overcome for two reasons. First, enterprise networks are already designed to log events at a population-scale. For example, our campus WiFi network generates 2 GB of syslog data comprising up to 11.5 million total events from approximately 58,000 devices and 5,500 APs on a typical weekday. Second, \wisleeps can use real-time location system (RTLS) reports the same way as syslog data. Specifically, reports of all devices by the RTLS are treated as association events. If a device reportedly disappears from an AP, it is treated as a disassociation event. \wisleeps can thus use either RTLS or syslog data equipped in existing WiFi networks, such as Cisco and Aruba \cite{Jaisinghani18}.

\subsection{Preprocessing Engine}
Our pre-processing engine takes in the syslog data (with anonymized MAC address) as input. Note, anonymization is performed on our campus IT department's server before data is copied to our system. The engine proceeds by partitioning event logs to construct per-device event logs of each user's primary device; the primary device is one that makes the largest number of daily AP associations (e.g., over a week). We maintain an up-to-date list of user devices to avoid pulling WiFi events from secondary and/or obsolete devices
(e.g., a user may change their smartphone to a new model). Finally, we apply a heuristic to identify devices with high activity presence in dorm areas as on-campus student residents. The pre-processing engine is written in python using 900 lines of code. 

\subsection{Change Point Detection Engine}
Processed per-device event logs are input for our detection engine. It computes WiFi event rates in 15 minutes time slots, spanning from 18:00 hours to 17:59 hours the next day. Our model predicts the sleep and wake-up time of users and delivers population-scale and individual-level analytics. We describe our model's performance results in Section \ref{sec:validation} and demonstrate our predictive analytics through several case studies in Section \ref{sec:cases}.

\subsubsection*{System Performance Metric:}
\label{sec:systemMetric}
Two performance measures are \textbf{\textit{accuracy}} and \textbf{\textit{timeliness}}. As reasoned in Section \ref{sec:approach}, our engine runs on an ensemble of models based on Bayesian change point detection to yield more acceptable accuracy despite working with weak priors. In Section \ref{sec:validation}, we present results from comparing the efficacy of \wisleeps compared to three baseline techniques (i.e., rule-based, normal and hierarchical priors) and tabulate the prediction accuracies in Table \ref{tab:metricAll}.

To achieve timeliness in delivering a population-scale analytics solution, our model utilizes Metropolis-Hashtings algorithm \cite{chib1995understanding}, which estimates the parameters $T_{sleep}$ and $T_{awake}$ for one user in approximately 4 seconds. We demonstrate in Section \ref{sec:scalability} how \wisleeps is computationally efficient in producing predictive analytics of 10,000 on-campus student residents under 12 hours.
While a single server is adequate to handle the processing needs on our campus, WiSleep uses a cluster to scale to larger user populations by parallelizing the analysis of user device traces across servers\footnote{Each server is a Dell PowerEdge R430 with 16 core 2.10 GHz Intel Xeon processor, 64GB RAM, 10 gigE network connections and local 1TB disk.}.
 In a practical use-case for our campus health administrators, \wisleeps can generate reports of sleep deprivation quickly enough to render pertinent insights.

\subsection{Analytics Platform}
\transhealth{Results from our unsupervised learning model extend to produce descriptive analytics of sleep patterns among large user groups. Specifically, it profiles anonymized users as regular or irregular sleepers based on their estimated sleep time, wake time, and sleep duration. It generates aggregated reports of these profiles at three timescales; day, week, and month to provide insight into anonymized profiles with aberrant sleep duration.} Section \ref{sec:cases} demonstrates several ways our data can be represented and how our findings support prior research on sleep studies, particularly on college students. Further, in Section \ref{sec:discussion}, we discuss how our analytics feature can be operationalized to several end-users for public health and personal use while upholding ethical considerations.

%% file: Validation.tex
\section{Experimental Evaluation}
\label{sec:validation}
We evaluate our model by first, assessing model performance from conducting a study among users living in campus dorms and private housing. Next, we compare model performance with other rule-based and Bayesian techniques.

\subsection{Datasets and User Study}
\label{sec:datauser}
\textbf{Ethical Considerations:} This paper's data collection and analysis were conducted under safeguards and restrictions approved by our Institutional Review Board (IRB) and Data Usage Agreement (DUA) with the campus network IT group. All device MAC addresses and authentication information are anonymized using a strong hashing algorithm. User identities were blinded by assigning numeric identifiers. Ground truth was collected within the IRB approved protocol. It is important to note that our population-scale analysis was performed on aggregate data of anonymous users. Individual analyses were performed on users who had consented to this study.\\

\noindent \wisleeps has been deployed on our campus and gathering event logs of all connected devices. Our university has over 31,000 students and close to 14,000 on-campus student residents. With approximately 58,000 detected devices, we anticipate 14,000 of these devices to be applicable for our sleep analyses.

\begin{table}[h]
\centering
\scalebox{1}{
\begin{tabular}{ |l|l|l| } \hline
\textbf{Study} & \textbf{Gender / Sleep Habit} & \textbf{Dataset} \\ \hline
\makecell[l]{Campus student residents: Fall 2021, 1 month \\In-home user: 2 weeks} & \makecell[l]{18M 2F (6R, 14IR)\\1M (1R)} & \makecell[l]{Identified WiFi network\\from events smartphone,\\Fitbit data, diary log}\\ \hline
Large-scale weekly analysis & 1000 student residents & Anonymized WiFi network  \\
 &  & events from smartphone \\ \hline
\end{tabular}}
\caption{\transhealth{Dataset summary. \emph{R} denotes regular sleeping habits, \emph{IR} denotes irregular sleeping habits.}}
\label{tab:userSummary}
\vspace{-4mm}
\end{table}
\vspace{-2mm}
\transhealth{Table \ref{tab:userSummary} summarizes our datasets. Data from our small-scale study is used for model validation. Our small-scale study is conducted in during the Fall 2021 on campus among 20 undergraduates and a single home-user participation. We precisely identified the participants' hashed MAC addresses by monitoring their WiFi events from a dedicated AP on-campus. Each student was given a Fitbit and kept diary logs for ground truth in all three phases. Simultaneously, we collected WiFi events of 1 homeowner for off-campus private housing validation. His event logs were collected from a home WiFi router. Separately, our case study consists of per-devices event logs of 1,000 students for a given day, demonstrating the types of sleep analytics \wisleeps can deliver on a large scale.}

\begin{table} [h!]
\centering
\scalebox{1}{
\begin{tabular}{|l |l|l|l|l|l|l| } \hline
\textbf{User-type}&\textbf{Parameter} & \textbf{Median} & \textbf{Mean} & \textbf{Max} & \textbf{Min} & \textbf{Stdev.} \\ \hline
\multirow{3}{*}{Regular}& Sleep-time & 01:45 AM & 02:00 AM & 10:00AM & 08:00PM & 02:45 \\
&Wake-up time & 09:45 AM & 10:18 AM & 02:00PM & 03:00AM & 02:40  \\
&Duration (hours) & 7.24 & 7.41 & 10.91 & 1.00 & 2.00  \\
 \hline
 \multirow{3}{*}{Irregular} &Sleep-time & 12:20 AM & 01:10 AM & 11:00 AM & 06:00 PM & 03:20  \\
&Wake-up time & 09:48 AM & 10:08 AM& 03:15 PM & 10:42PM & 02:40\\
&Duration (hours) & 7.10 & 6.80 & 10.91 & 0.63 & 1.95  \\
 \hline
\end{tabular}}
\caption{\transhealth{Sleep ground truth summary for campus student residents.}}
\label{tab:sleepsummary}
\end{table}

\transhealth{Table \ref{tab:sleepsummary} summarizes our participants' sleep logs. We adapted the consistency metric proposed by Rashid \etals \cite{rashid2017revisiting} to generate a sleep consistency score between 0 to 1 for each user — 1 denotes the user as having regular sleep patterns throughout the week. We applied a median split to determine the threshold for categorizing users into groups with regular sleep patterns (score 0.75 to 1) and irregular sleep patterns (0 to 0.75). As per Table \ref{tab:userSummary}, we identified more users with irregular sleeping patterns than regular. While we observed users receiving comparable sleep duration between the different sleep regularity profiles, on the whole, keeping a regular sleep schedule is advantageous for achieving good sleep quality \cite{brown2002relationship}.}

\subsection{Validation Study}

The first validates our approach and utilizes ground truth data from the user study dataset. We compare our prediction values, $T_{sleep}$ and $T_{awake}$, with the ground truth Fitbit data and compute four metrics: Accuracy, Precision, Recall, and F-score. Accuracy is the proportion of correct predictions (both sleeping or awake periods) relative to all predicted sleeping or awake periods. Precision is the ratio of all correct sleep/awake periods to the total number of predicted sleep/awake periods. F-score indicates the optimal balance that maximizes precision and recall (a score of 1 indicating a perfect predictor). 

\begin{table} [h!]
\centering
\scalebox{1}{
\begin{tabular}{ |l|c|c|c|c|c| } \hline
\textbf{} & \textbf{Accuracy} & \textbf{Precision} & \textbf{Recall} & \textbf{Fscore} & \textbf{Sleep,wake-up time error}\\\hline
Campus residence & 91.98\% & 80.29\% &	91.06\% &	0.90 & 51, 36 minutes\\ \hline
In-home & 85.00\%  &	76.49\%  &	82.89\%  &	0.79  & 70, 42 minutes\\ \hline
Average & 88.50\%  & 78.39\%  &	86.90\%  & 0.84  & 60, 39 minutes\\ \hline
\end{tabular}}
\caption{\transhealth{\wisleep's performance for different study environment.}}
\label{tab:wisleepUserPerformance}
\end{table}

\transhealth{Table \ref{tab:wisleepUserPerformance} summarizes our final results (after performing error analysis, explained in Section \ref{sec:comparisonbaseline}) with \wisleeps achieving an average accuracy of 88.50\% compared to Fitbit ground truth  (+/- 4.70\%, max: 96.00\%, min: 79.25\%) for predicting users' sleep in campus residence. In a private home-use, \wisleeps yields 85.00\% (+/- 2.91\%, max: 93.00\%, min: 81.00\%). As \wisleep's average performance varies between $\approx$ 85\%-92\%, we seek to understand the cases in which our technique performs and breaks. Overall, our system produces error within 60 minutes for estimating sleep time, and 39 minutes for wake-up time.}

\begin{figure*}[h]
    \centering
    \begin{tabular}{cc}
    \includegraphics[width=0.45\linewidth]{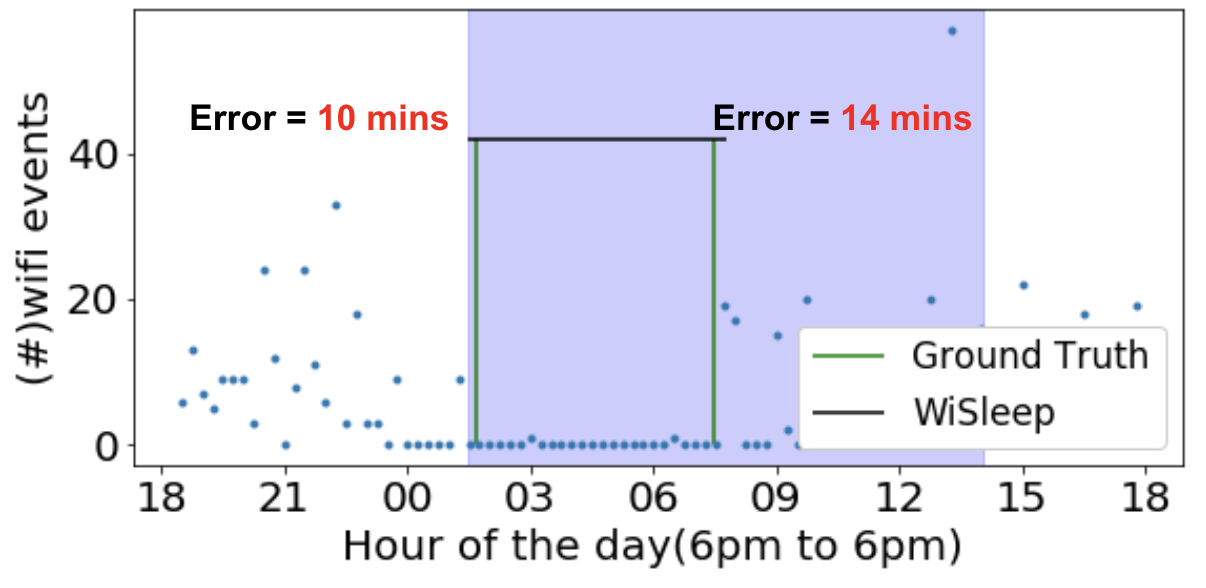} &
    \includegraphics[width=0.45\linewidth]{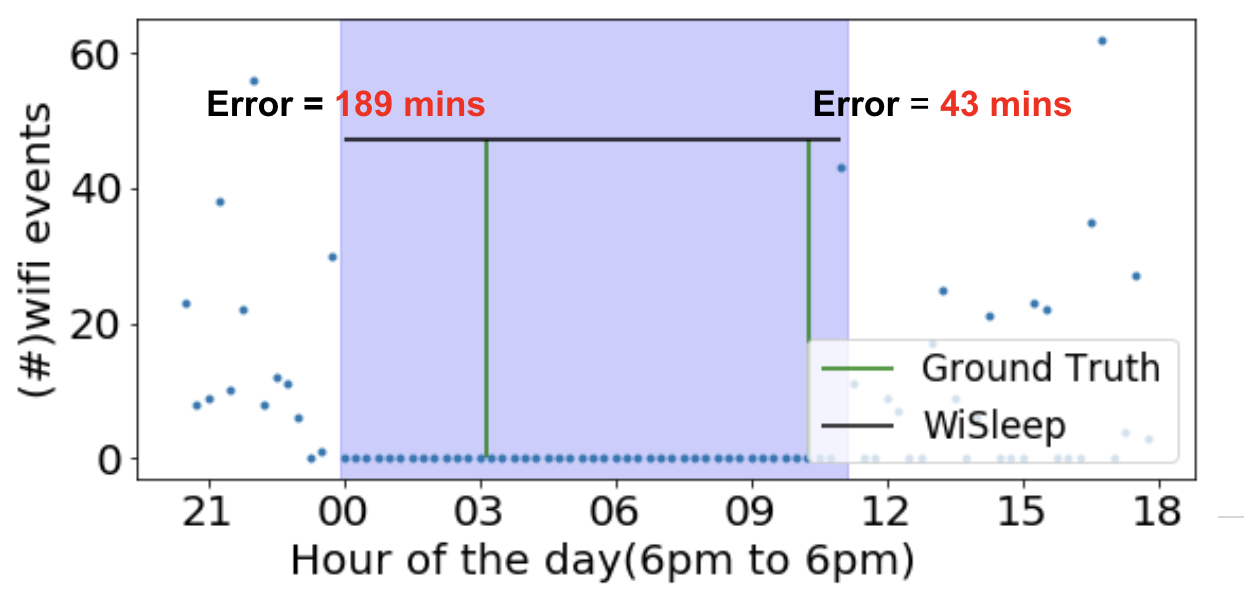} \\
    (a) Best case & (b) Worst case \\
    \end{tabular}
    \caption{\transhealth{\wisleep's performance for a user, $P13$, living on campus. Shaded area denotes WiFi events in the residential area.}}
    \label{fig:userWifi}
\end{figure*}

\transhealth{Figure \ref{fig:userWifi} illustrates the WiFi event traces and the predicted/actual sleep/wake-up times of one student resident, $P13$. Figure \ref{fig:userWifi} (a) exemplifies the best case prediction, with our model yielding 95.84\% accuracy (10 minutes sleep time error, 15 minutes wake-up time error, 06 hours 35 minutes sleep duration). In this case, as with most days, the user exhibited near-ideal behavior where sleep onset occurs shortly after ceasing phone activity. Our sleep duration estimation for $P13$ supports prior reporting that the majority of young adults use mobile phones happens at least one hour prior to sleep \cite{gradisar2013sleep} and at least within 10 minutes soon after they wake up \cite{newyorkpost}.}

\transhealth{In contrast, Figure \ref{fig:userWifi} (b) illustrates $P13$'s erroneous prediction, with the model performing at 79.25\% accuracy (3 hours 52 minutes sleep duration). Specifically, $ P13$'s WiFi event traces did not match his sleep time by more than 189 minutes and wake-up time by 43 minutes, despite the system acquiring no WiFi events between midnight to 11:00 am the next day. As previously discussed in Section \ref{sec:designRationale}, deviations in user behaviors can significantly affect system accuracy (e.g., for example, sleep onset latency and delayed use in the smartphone). While the true reasons behind these deviations are undetermined, deviations of user behaviors will pose a limitation. However, these inaccuracies can efficiently be detected as an anomaly.}


\subsubsection*{Private Home Use}

To demonstrate the applicability of our system in a private setting, we tested \wisleeps with one home user and one WiFi AP deployed. In a typical home network setup, a user will have the option to set their mean sleep and wake-up times as part of initializing \wisleep. For our user, these times were set to 11:00 pm and 7:00 am, respectively. It is important to note that in this study, we are only tracking a single home user. We discuss in Section \ref{sec:discussion} how multiple home users can be monitored by \wisleep.

\begin{figure} [h!]
	\centering
	\begin{tabular}{cc}
	\includegraphics[width=.4\textwidth]{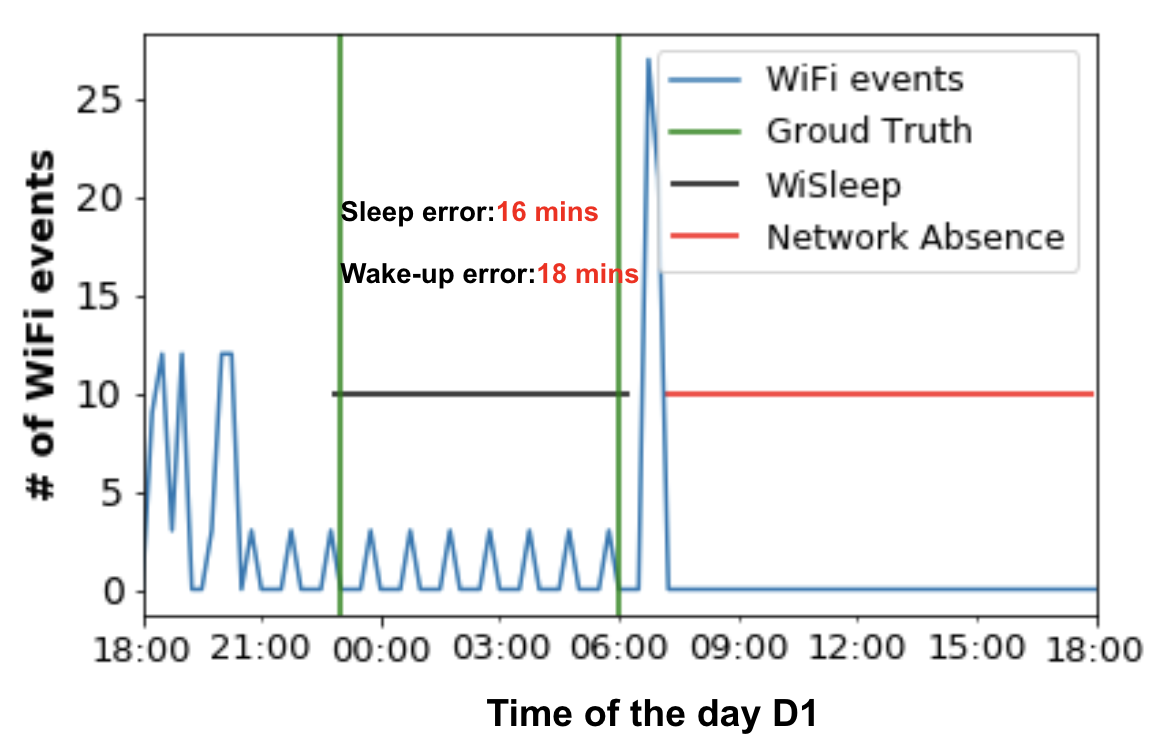}&
	\includegraphics[width=.4\textwidth]{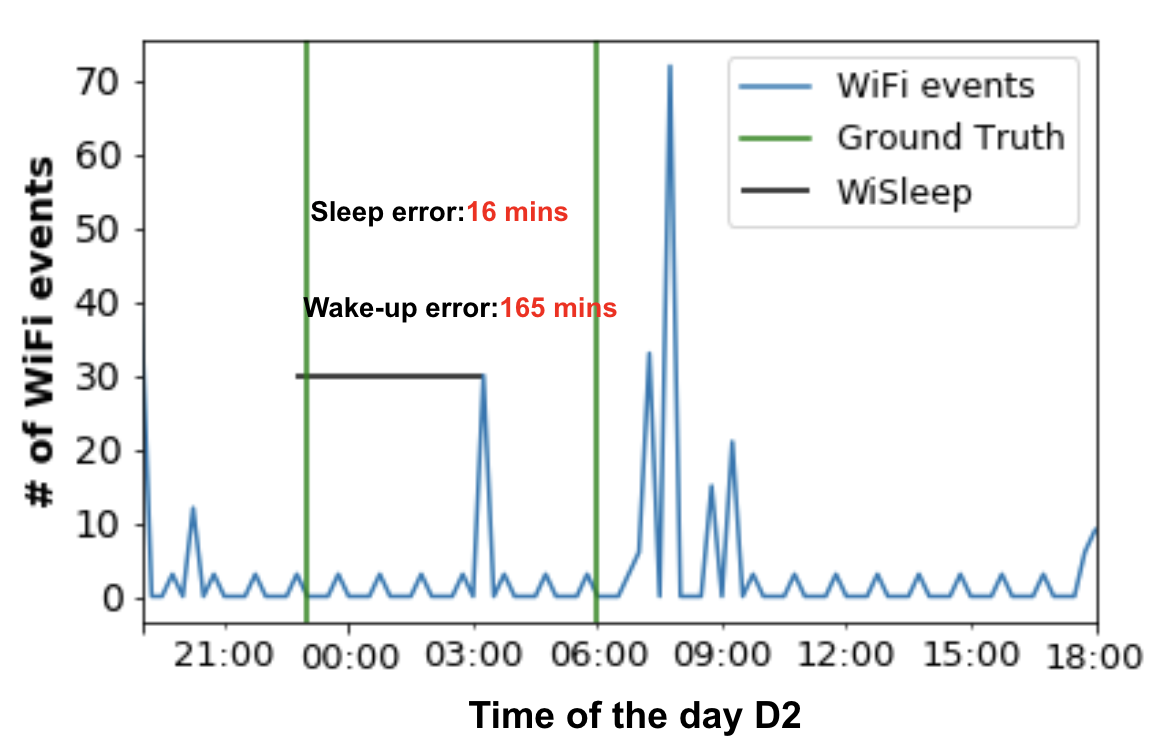} \\
	(a)  Best case & (b) Worst case \\
	\end{tabular}
	\caption{\wisleep's  performance for a user in a home network.}
	\label{fig:homewifi}
\end{figure}

\transhealth{
Upon running our model for two weeks, \wisleeps successfully yields approximately 85\% accuracy  (70 minutes average sleep time error, 42 minutes average wake-up time error, 07 hours 05 minutes average sleep duration). Figure \ref{fig:homewifi} charts the number of WiFi events detected from the user's primary device, and accordingly, his predicted sleep duration and ground truth for two different days, $D_{1}$ and $D_{2}$. $D_{1}$ illustrates an essential aspect of our implementation that does not falsely predict sleep as a result of network absence. On $D_{1}$, the user was confirmed to not be present at home between 8:00 am and 6:00 pm. For this reason, the WiFi network captured no network activity, including periodic pings, which would otherwise be recorded had the user (and his primary device) been physically present.
In comparison, $D_{2}$ represents a day where \wisleeps underestimates the user's sleep duration. Here, our model incorrectly inferred the user as waking up from detecting high network activity at 03:15 am, illustrating a deviation in device behavior. We investigate in the next section the breaking point of our model to background activities in Section \ref{sec:noise2}. Additionally, a potential solution is employing strong priors in the ensemble model to overcome this device-specific challenge, which would work especially well for users with regular sleeping habits.}

\subsection{Comparison with State-of-the-art}
\label{sec:comparisonbaseline}

\transhealth{Next, we conduct an in-depth analysis by comparing \wisleeps with prior learning techniques of predicting sleep}: a rule-based heuristic and two state-of-the-art Bayesian methods. Our rule-based heuristic first determines a user's residential dorm. It classifies the time (slot) spent in their dorm as active or inactive by checking if the observed WiFi rate $>$ 2 (2 is chosen to ignore the periodic pings). Accordingly, the longest inactivity interval is decided as the user's sleeping period. The second is a Bayesian approach using Normal priors, as proposed by El-Khadiri \etals \cite{el2018sleep}. The third is also a Bayesian approach using hierarchical priors, as proposed by Cuttone \etals \cite{cuttone2017sensiblesleep}.

\begin{table}[h!]
\centering
\scalebox{.8}{
\begin{tabular}{r|p{1.5cm}|p{1.5cm}|p{1.5cm}|p{1.5cm}|p{1.5cm}|p{1.5cm}|p{1.5cm}|p{1.5cm}}\hline
\textbf{} & \multicolumn{2}{c|}{\textbf{Ensemble}} & \multicolumn{2}{c}{\textbf{Hierarchical}}  & \multicolumn{2}{c}{\textbf{Normal}}& \multicolumn{2}{c}{\textbf{Rule-based}}\\
 & $T_{sleep}$ & $T_{Wake}$ & $T_{sleep}$ & $T_{wake}$ & $T_{sleep}$ & $T_{Wake}$ & $T_{sleep}$ & $T_{Wake}$ \\ \hline
\textbf{Median} & 62 & 35 & 77 & 50 & 93 & 37&270 &147 \\
\textbf{Mean} & 77 & 55 & 110 & 78 & 122 & 57 &240 &192 \\
\textbf{Max} & 192 & 169 & 234 & 170 & 210 & 164 &640&387\\
\textbf{Min} & 0 & 0 & 4 & 1 & 1 & 0 &1 &1  \\

\textbf{Stdev.} & 70 & 59 & 89 & 72 & 103 & 49 &150  &136\\
\textbf{Q1, Q3} & 24, 80 & 11, 65  & 26, 92 & 25, 79 & 47, 87 & 12, 61  &140, 320 &70, 190\\
\textbf{UIF, UOF} & 163, 272 & 146, 227  & 191, 290 & 160, 241 & 147, 207 & 134.5, 208 & 590, 860 & 370, 550\\
\hline
\end{tabular} }     
\caption{\transhealth{Summary statistics of sleep time error and wake-up time error in minutes. } }
\label{tab:summaryEstimationError}
\vspace{-4mm}
\end{table}

\transhealth{Table \ref{tab:summaryEstimationError} summarizes the descriptive statistics predicting sleep and wake-up times across all methods compared with Fitbit ground truth. Overall, \wisleeps yields an average error of 77 minutes in sleep time and 55 minutes in wake-up time (\wisleep: 88.50\%, Normal: 83.90\%, Hierarchical: 85.20\%, Rule-based: 72.10\%). The data distribution reveals that a small number of outlier days contribute to a large fraction of the inaccuracies. We calculated the upper inner fence (UIF), the third quartile plus 1.5*IQR, and the upper outer fence (UOF), the third quartile plus 3*IQR, as thresholds for exclusion. Removing outliers resulted in excluding 66 days of prediction among 18 participants, including the examples we discussed in Figures \ref{fig:userWifi} (b) and \ref{fig:homewifi} (b).}

\begin{table} [h!]
\centering
\scalebox{1}{
\begin{tabular}{ |l|c|c|c|c|c|c| } \hline
\textbf{} & \textbf{Accuracy} & \textbf{Precision} & \textbf{Recall} & \textbf{Fscore} & \textbf{Sleep,wake-up time error}&
\textbf{p value}\\\hline
\wisleep & 88.50\%  & 78.39\%  & 86.90\%  &   0.84 & 60, 39 minutes & - \\
Normal\cite{el2018sleep} & 83.90\% & 70.71\% & 93.10\% & 0.81 & 115, 50 minutes & .01\\
Hierarchical\cite{cuttone2017sensiblesleep}  & 85.20\% & 73.08\% & 88.32\% & 0.84 & 102, 59 minutes & .01\\
Rule-based & 72.10\% & 50.23\% & 15.83\% & 0.46 & 187, 190 minutes & .001\\\hline
\end{tabular}}
\caption{\transhealth{\wisleep's performance compared against three baselines.}}
\label{tab:metricAll}
\vspace{-5mm}
\end{table}

\transhealth{Table \ref{tab:metricAll} summarizes our model performance of all techniques compared with Fitbit ground truth. \wisleeps significantly outperforms the rule-based technique by over 16\% accuracy at $p<.001$. While accuracy improvement is within 3-4\% compared with other Bayesian techniques, \wisleeps effectively reduces sleep time error within 60 minutes and wake-up time error within 39 minutes. The differences in errors are significantly lesser ($p<.01$) than Normal and Hierarchical Bayesian. Given that our dataset consists of more irregular sleepers than regulars (see Table \ref{tab:userSummary}), the uniform priors, along with location information our ensemble technique employs, will be able to capture different sleeping habits giving weights to appropriate priors.}

\begin{figure*}[h]
    \centering
    \begin{tabular}{cc}
    \includegraphics[width=0.45\linewidth]{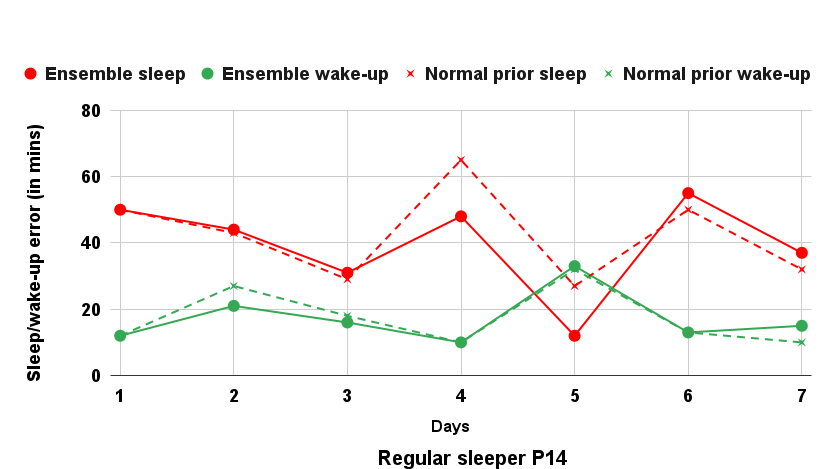} &
    \includegraphics[width=0.45\linewidth]{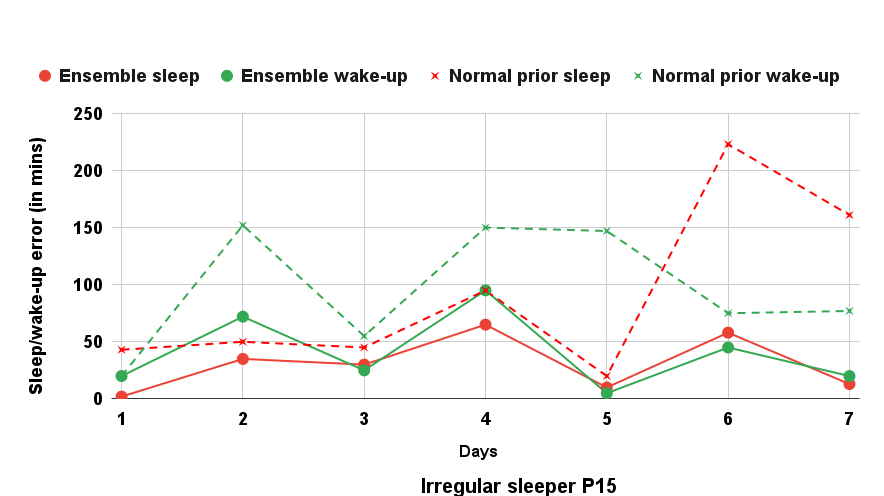} \\
    \end{tabular}
    \caption{\transhealth{\wisleep's ensemble versus normal prior for two students with regular sleep (left) and irregular sleep (right) patterns. }}
    \label{fig:regularity}
\end{figure*}

\transhealth{To illustrate the benefits of our ensemble approach, let us consider two users, $P14$ (a regular sleeper) and $P15$ (an irregular sleeper), as per Figure \ref{fig:regularity}. For $P14$, the ensemble model and normal priors achieve comparable performances overall. In accommodating users with irregular sleep patterns, as per $P15$, the model employing only normal priors would produce larger errors, especially in predicting wake-up times. On days 6 and 7, where sleep time significantly deviates from the population norm, errors from using normal priors were beyond 150 minutes.}

%% file: Scalability.tex
\section{Practical Considerations}
\label{sec:practical_considerations}
\transhealth{Here, we investigate \wisleep's robustness to device-specific challenges listed in Section \ref{sec:designRationale}. Additionally, our experiment trials the system to predict an extensive group of users to support large-scale analytics.}

\subsection{Noisy Data: Ping-Pong Effects}
\label{sec:noise1}

\transhealth{When a stationary device is within the range of connecting to several APs with similar signal strengths, it may connect with one AP. However, the connection can also switch back and forth between different APs, causing a spectrum handoff known as the ``ping-pong'' effect.} The noise from this effect can resemble network activity despite the absence of user interactions. \transhealth{To avoid errors as a result of ping-pong effects}, we group APs in an area, such as a dorm floor, and filter out patterns that resemble ping-pongs between nearby APs. 

\begin{figure}[h!]
    \centering
    \includegraphics[width=0.6\linewidth]{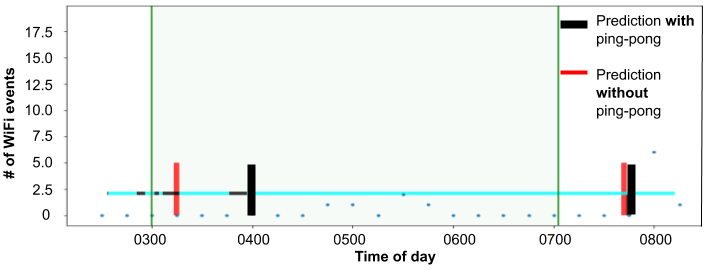}
    \caption{Ping pong events during P11's sleep period.  Green shaded area denotes the ground truth. Cyan horizontal line denotes the primary AP that user is usually connected to, and black horizontal lines denote other APs in close proximity.}
    \label{fig:userpingpong}
\end{figure}

Consider user P11, whose phone exhibits significant ping pong noise, as shown in Figure \ref{fig:userpingpong}. We observe multiple ping-ponging events between 3:00 am - 4:00 am, but the connection remains consistent with the primary AP throughout the rest of the sleep period. Without our heuristic, \wisleeps would have predicted sleep only after the connection stabilizes at 4:00 am, with a 45 minutes delay from the actual sleep time (instead of 3:00 am, closer to the actual value).

\subsection{Noise from Background Activities}
\label{sec:noise2}

\begin{figure}[h!]
    \centering
    \includegraphics[width=0.6\linewidth]{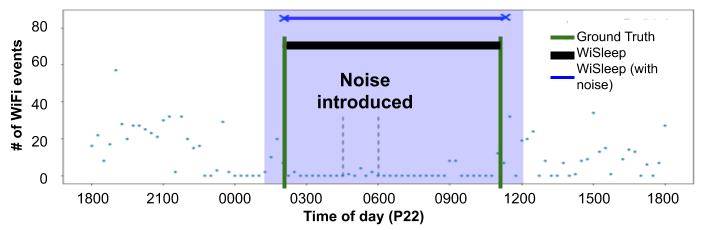}
    \caption{\wisleep's performance with modest background activities as noise. Shaded area denotes residential area.}
   \label{fig:scale_background}
\end{figure}

Background activities such as push notifications and software updates can introduce noise by appearing to have active network usage when a user is asleep. To evaluate the impact of such noise, we carefully introduced background activities in a controlled fashion. Here, we used an Android phone, alternating between long periods of idle, followed by some push notifications, and finally, a mobile app download from the Play store. We created a synthetic device trace by inserting noisy traces into an actual device during a nightly sleep period. The synthetic trace, shown in Figure \ref{fig:scale_background}, was then subjected to our change point detection method. As shown, the sleep and wake-up times before and after the noise injection are quite similar ($\approx$ 15 minutes difference in wake-up time). 

\transhealth{\wisleeps is adaptive to a modest amount of noise from background activities. Since the frequency of push notifications and app updates are typically low, our ensemble method is resilient to noise introduced by their presence. Unfortunately, this also indicates that our technique will produce inaccuracies when high amounts of noise events are present. As discussed in Figure \ref{fig:homewifi} (b), a potential solution is employing strong priors in the ensemble model. However, if a user has the habit of streaming movies (which may result in an event peak) at the start of their sleep, the model is likely to predict delayed sleep behavior.}

\subsection{Impact of Inactive Periods}

\begin{figure}[h]
	\centering
	\includegraphics[width=.45\textwidth]{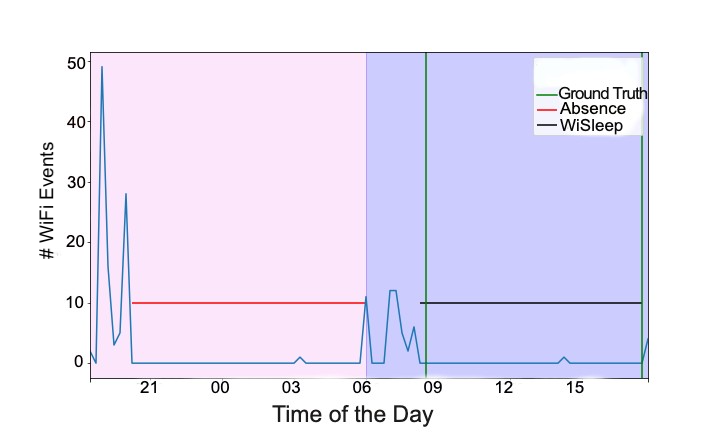}
	\caption{Impact of inactive periods on \wisleep's performance in non-residential (red shade) and residential (blue shade) areas.}
	\label{fig:falsepositive}
\end{figure}

There can be many device inactivity periods for a user in a day, leading to false positives. \wisleeps accommodates false positives by picking only the relevant inactive periods using priors for sleep and wake-up times and then considering a user's physical presence in their residential area. For instance, in Figure \ref{fig:falsepositive}, we observe that a user was inactive at two time periods; first, 8:15 pm to 6:00 am and second, from 8:15 am to 5:30 pm. Inactivity between 8:15 pm to 6:00 am is typically classified as the eventual sleeping duration, primarily because the user is in residence. However, this example illustrates a different case -- the user is in residence between 8:15 am to 5:30 pm (this is highlighted in light blue area). A similar situation can be seen in Figure \ref{fig:homewifi} (b), where \wisleeps can identify network absence, thus avoiding false positives.

\subsection{System Scalability}
\label{sec:scalability}

 In a real-world implementation of a sleep analytics solution for on-campus student residents, \wisleeps needs to scale to tens of thousands of users present on campus. Next, we evaluate the scalability of the \wisleeps system to support a large number of users under accuracy and timeliness constraints. To validate our argument, we examine two factors -- 1) the number of samples needed for computation and 2) the CPU cost of the sampling process. First, we determine the number of samples needed for each user to create accurate estimates in the sampling process employed by \wisleeps. Generally, the more samples used, the higher the accuracy. However, we must also consider that higher samples will result in higher CPU cost, affecting the results' timeliness.

\begin{figure}[h!]
    \centering
    \includegraphics[width=0.6\linewidth]{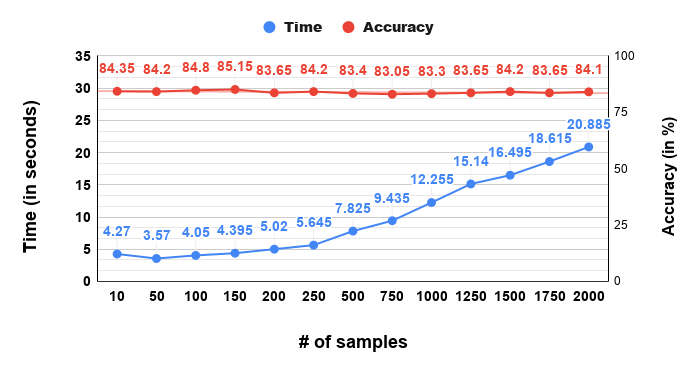}
    \caption{Accuracy and CPU overhead of change point detection for various sample size.}
    \label{fig:scale_regular}
\end{figure}

Figure \ref{fig:scale_regular} shows the accuracy and the CPU cost of the computation for two different users obtained by varying the number of samples from 10 to 2000 over one week. We observe that using between 10 to 50 samples yields an accuracy of approximately 85\%, which does not significantly change as the sample size increases. Naturally, the more samples used, the higher the CPU cost. The results show that a good accuracy -- computation trade-off for \wisleeps is to use 50 samples producing an accuracy of 85\% with a CPU processing cost of approximately 4 seconds per user.

\begin{figure*}[h]
    \centering
    \begin{tabular}{cc}
    \includegraphics[width=0.4\textwidth]{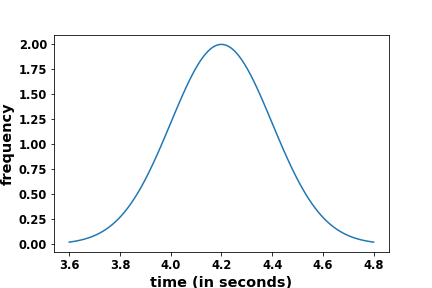} &
    \includegraphics[width=0.44\textwidth]{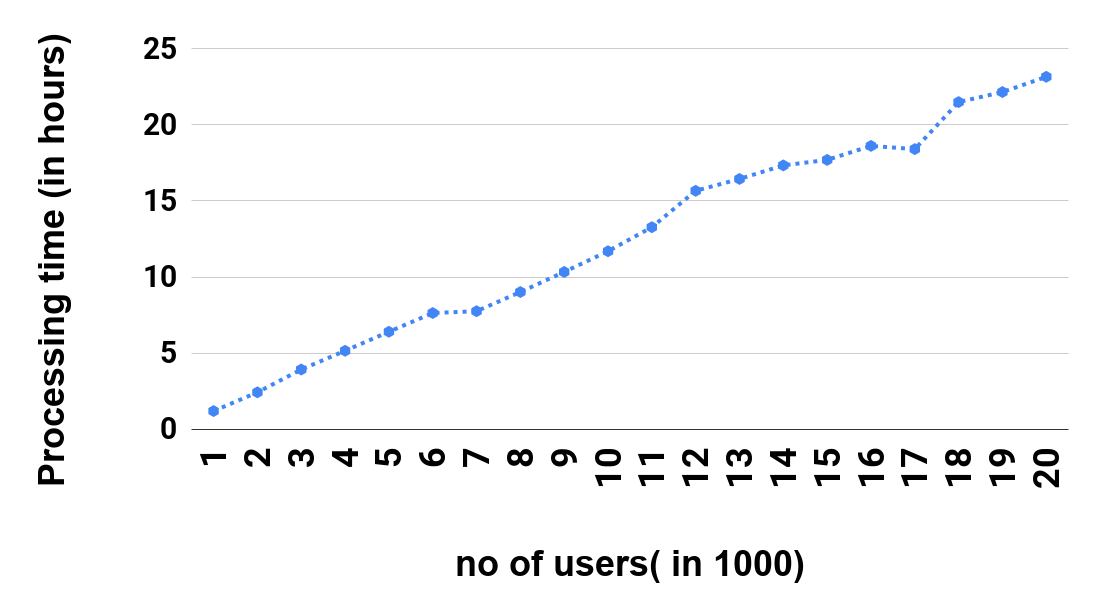} \\
    (a) distribution of time taken per user & (b) CPU overheads for 1,000 to 20,000 users \\
    \end{tabular}
    \caption{WiSleep scales up to $>$ 20k users on a single server.}
    \label{fig:scale_users}
\end{figure*}

Next, we examine how \wisleeps scales when processing many users. Figure \ref{fig:scale_users} shows that the CPU time scales linearly with the number of users, and the prediction cycle is completed in  23 hours for 20,000 users,  thus showing that a single server is sufficient to handle all on-campus students at our university. Hence, \wisleeps can generate reports of sleep deprivation of a large number of users quickly enough to render pertinent insights on the same day. One key point is that our system currently uses unoptimized python libraries for Bayesian inference and does not use any hardware accelerators such as GPUs. Additionally, the computation is highly parallelizable and can be scaled near-linearly by using a cluster of servers.

%% file: Casestudy.tex
\section{\wisleeps Analytics}
\label{sec:cases}
We present insights from \revise{our large-scale on-campus study two cases demonstrating} how our population-scale aggregate analytics can benefit public health and personal use.

\subsection{Population-scale Aggregate Analytics}
\label{sec:case-oncampuspopulation}

\begin{figure}[h!]
    \centering 
    \includegraphics[width=.5\linewidth]{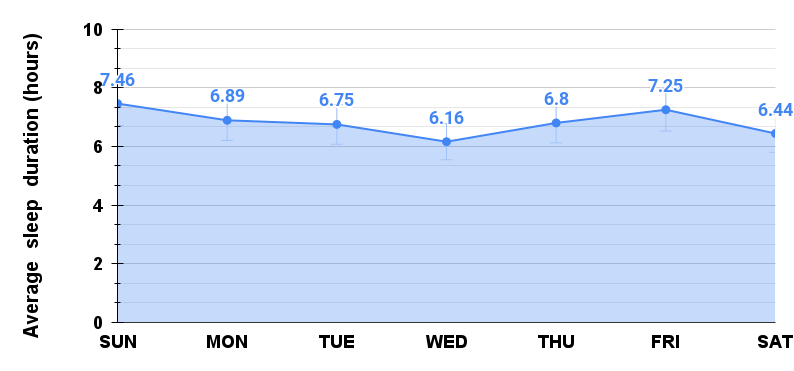}
    \caption{\transhealth{How do aggregated sleep patterns vary by day of the week? Mean sleep duration predicted for 1000 users.}}
    \label{fig:population_sleep}
\end{figure}

Using our large-scale dataset of 1000 anonymous student users, we conduct an aggregate-level analysis of their sleep behavior for one week. Figure \ref{fig:population_sleep} plots the average sleep duration of all users by the day. Overall, the results support existing findings that students sleep between 6 to 7 hours, and longer on Sundays \cite{buboltz2001sleep,finlay2012leisure}. \transhealth{We recognize a slight declining sleep trend at the beginning of the week, before gradual increments later in the week. We expect the decrease in sleep on weekdays is likely due to students fulfilling various academic demands, while Saturday could be attributed to more active extra-curricular activities \cite{finlay2012leisure}.}

\begin{figure}[h!]
\centering
    \centering
    \includegraphics[width=0.4\linewidth]{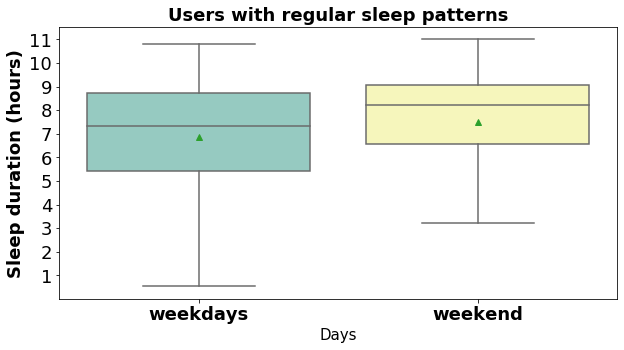}
    \includegraphics[width=0.4\linewidth]{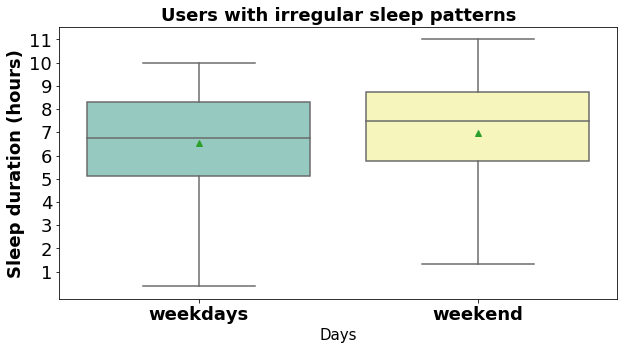}
\caption{\transhealth{Box plots comparing the predicted sleep duration difference between users with regular and irregular sleep patterns on weekdays and weekends. Green triangle indicates the mean.}}
    \label{fig:population_regular}
\end{figure}

\transhealth{To better understand these results, we find that 556 students out of our 1000 students ($>50\%$) exhibited irregular sleep patterns. This percentage is similar to our small-scale sample as summarized in Table \ref{tab:userSummary}. We compare the sleep duration over weekdays and weekends between these sleep profiles, as shown in Figure \ref{fig:population_regular}. We observed that, on average, irregular sleepers receive approximately 6.5 hours of sleep, comparable to regular sleepers who get approximately 7 hours of sleep. The aggregated analysis from our large-scale prediction aligns with the ground truth results of our small-scale study (see Table \ref{tab:sleepsummary}).}

\subsection{Individual-level Sleep Analytics}
\label{sec:case-oncampusindividual}
Next, we illustrate \wisleep's ability to perform sleep analytics for individual on-campus student users for a semester. We selected two users \revise{from our campus user study} and retrieved WiFi events for approximately 70 days from the start of the semester till the semester-end. Note that we intentionally left out the first three weeks, as students were more likely to take this time still to settle into their student accommodation.

\begin{figure}[h!]
    \centering
    \includegraphics[width=0.65\linewidth]{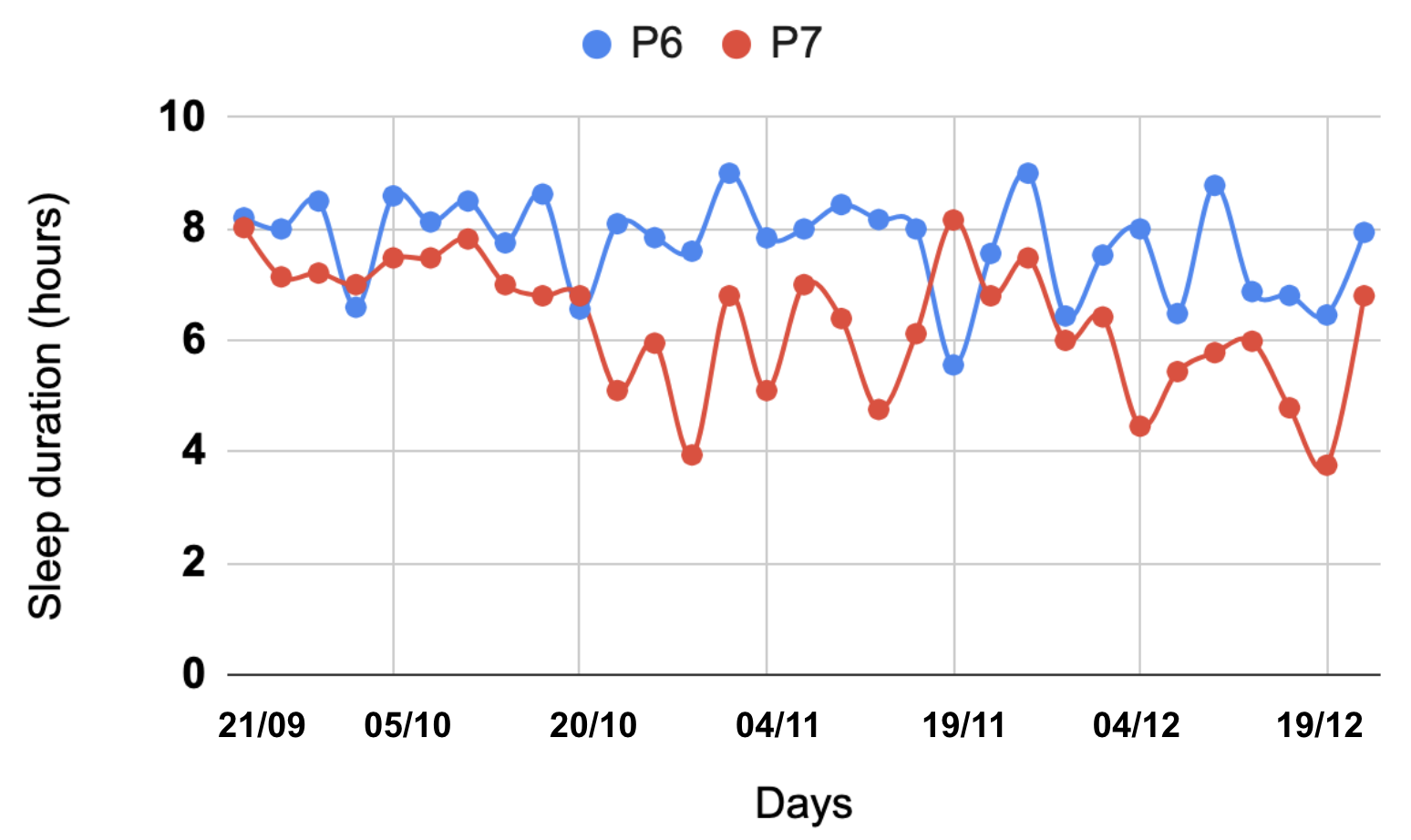}
    \caption{\transhealth{How do sleep patterns change over a semester? Predicted sleep duration for two participants, $P6$ and $P7$.}}
    \label{fig:longitudinal_sleep}
\end{figure}

Figure \ref{fig:longitudinal_sleep} illustrates the predicted sleep duration, averaged every three days for two users, $P6$ and $P7$. \transhealth{On the whole, both users display sleep inconsistencies throughout the semester. However, $P7$'s sleep patterns seemed fairly consistent at the start of the semester and showed high variability as they transitioned to mid-term week (20/10) until the end of the semester (19/12).}

\begin{figure}[h!]
    \centering
    \begin{tabular}{cc}
    \includegraphics[width=0.4\linewidth]{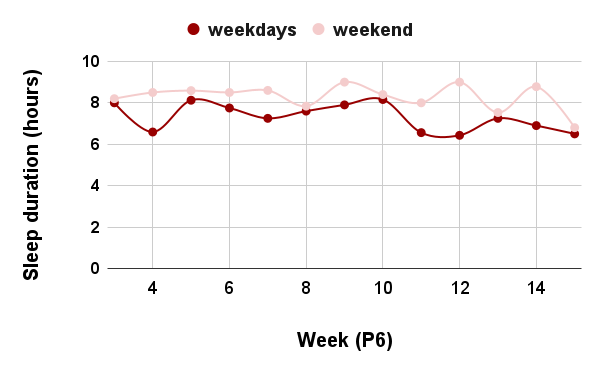} &
    \includegraphics[width=0.4\linewidth]{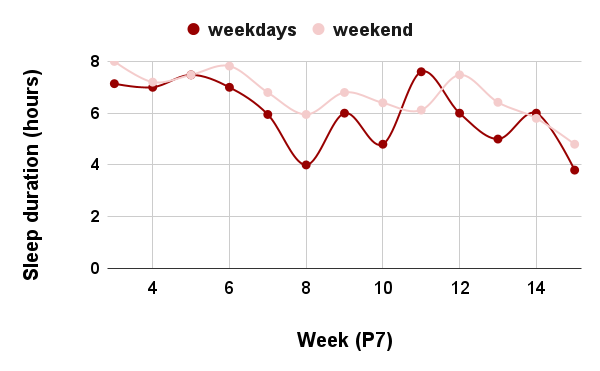}\\
    (a) Regular sleeper & (b) Irregular sleeper\\
    \end{tabular}
    \caption{ \transhealth{Inferred sleep duration for 2 participants, $P6$  and $P7$ between weekends and weekdays, over the semester.}}
    \label{fig:longitudinal_weekbreakdown}
\end{figure}

\transhealth{Figure \ref{fig:longitudinal_weekbreakdown} illustrates the sleep regularity pattern for the two users on weekdays and weekends each week over the semester. In general, both users received more sleep on weekends than on weekdays; $P6$ weekdays: average sleep =  07 hours 40 minutes, std = 01 hours 20 minutes, weekends: average sleep = 08 hours 05 minutes, std = 00 hours 45 minutes. $P7$ weekdays: average sleep = 6 hours 00 minutes, std = 01 hours 10 minutes, weekends: average sleep = 6 hours 40 minutes, std = 00 hours 50 minutes. However, on Week 8, corresponding to the mid-terms, we observed $P7$ receiving approximately 4 hours of sleep on the weekdays. This trend persisted throughout weeks 10 and 13 and reached the lowest in Week 15. On weekends, his sleep pattern showed a consistently declining trend from Week 12 onwards.}

\transhealth{Indeed, students sleeping less towards the end of the semester are typical. However, users who exhibit a considerable reduction in their sleep throughout extended periods can be a cause of concern. In the following section, we discuss how our system can contribute to opportunities for promoting sleep health.} 

%% file: Discussion.tex
\section{Discussion}
\label{sec:discussion}
\transhealth{We have addressed the challenges of estimating sleep duration and providing analytics for population and individual use. Here, we discuss the implication of \wisleeps and its limitations.}

\subsection{Opportunities for Public Health \transhealth{and Personal Well-being}}
Indeed, poor sleep hygiene has major health consequences and is a public health issue \cite{perry2013raising,altevogt2006sleep,adolescent2014school}. \wisleeps plays a vital role in responding to the call-for-action to advance sleep disorder problems at the forefront of public health. Among the key strategies promoted, Perry \etals raised the need for improving sleep surveillance among the younger population \cite{perry2013raising}. In an ideal scenario to strengthen universities' health and well-being services, \wisleeps can help health professionals decide on critical times to raise community-level awareness on healthy sleep behavior among students. Beyond that, \wisleeps aggregated analytics can provide ``open-access'' data sources for public health researchers to study sleep disorders properly. Often, sleep screening is excluded in health care screenings compared to those of eating and drinking behaviors \cite{sorscher2008your}. Thus health professionals do not have enough knowledge of sleep disorders \cite{papp2002knowledge}. 

\transhealth{As stated in Section \ref{sec:datauser}, our approach requires access permission on the institutional level, in this case, from the IT department of our university. In practice, the device information is hashed to maintain user anonymity. We had only identified devices of users who consented to the study. Collecting WiFi network event logs does not consist of critical and sensitive information such as browsing content and apps used. This privacy-preserving data modality can allow for sleep analytics without violating data-sharing practices for public health \cite{parker2015sharing}.}

\transhealth{Deploying interventions remains an option for students who struggle to recover from their lack of sleep, as observed for $P7$ in Section \ref{sec:case-oncampusindividual}, and who prefers seeking professional help. To mobilize personal sleep health services, users must consent to be notified through their devices. We imagine this process will require disclosing their student identification and MAC address of the primary device to progress to a professional health and well-being counselor. Similarly, private homeowners can allow the system to collect WiFi network events logs from their home AP. In a multiple-user home setting, initialization linking users to their devices must be established before monitoring takes place.}

\subsection{Extension to Detect Polyphasic Sleep}
An essential extension is inferring polyphasic sleep. Prior research suggests people who generally sleep $< 5$ hours during the night are more likely to sleep in daytime \cite{liu2000sleep} and those who sleep  $> 1$ hour at daytime are more likely to sleep less at night time \cite{ancoli2006insomnia}. We could conceivably apply a set of rules to check for secondary sleep on days users are monitored to have slept for $< 5$ hours, either before the sleep time ($T_{sleep-1}$) or after the wake-up time ($T_{awake+1}$). Our model could use only a uniform prior (see Section \ref{sec:ensembleModel}) to find the longest inactive period at these two times. For instance, in Figure \ref{fig:falsepositive}, if both the inactive periods had occurred in a residential building, the first or second inactive period could be classified as `secondary sleep.' This warrants further investigation.

\subsection{Limitations}
First, our approach assumes that device event data is available on a longitudinal basis for \transhealth{estimating daily sleep duration.} However, data for students may be absent from our logs for numerous reasons. Data unavailability will disrupt \wisleeps from its daily monitoring. Our approach is also not free from periodic maintenance to ensure user devices are valid (e.g., users did not change their phones) and maintain current residence (e.g., students may no longer reside in dorms). \transhealth{Also, a high amount of background noise can sometimes lead to inaccurate predictions, primarily when these noises are attributed to user-specific challenges, such as sleep onset latency or insomnia.} Finally, it is essential to emphasize that the goal of our approach is to infer sleep duration of users -- and not detect the nuances in sleep characteristics  \cite{rahman2015dopplesleep,Nguyen16} that prior work is set to achieve. \transhealth{The utilization of coarse-grained data inherently limits our ability to determine sleep onset latency, differentiate sleep stages, detect sleep apnea and insomnia. There is no workaround to this limitation, but our approach remains relevant to achieve the larger goal of complementing population-scale analytics by estimating users' sleep and identifying those with aberrant sleep duration}, to which coarse-grained WiFi information is more than adequate.

%% file: Conclusions.tex
\section{Conclusions}
\label{sec:conclusion}

In this paper, we presented \wisleep, a network-based system to detect sleep periods by passively observing the network activity of a user's phone and provide aggregated and individual-level analytics to accommodate for public and personal use. We presented an ensemble-based Bayesian inference technique to infer sleep from coarse-grain WiFi association and disassociation events. We validated our approach using \transhealth{20} users living on campus dormitories and 1 private homeowner in our study. We showed that \wisleeps outperforms the state-of-the-art methods for users with irregular sleep patterns while \transhealth{yielding comparable accuracy of 88.50\% within 60 and 39 minutes of sleep and wake-up time errors}. Further, we showed that \wisleeps can process the data of 20k users on a single commodity machine, allowing it to scale to large campuses with low server requirements. Our large scale case study revealed several interesting insights for population-scale and individual sleep analytics. As future work, we plan to combine our sleep inference model with stress detection methods to develop a complete student well-being service.